\documentclass[a4paper,twoside]{article}
\newcommand{\affil}[1]{$^{\rm #1}$}
\usepackage[authoryear]{natbib}
\bibpunct{(}{)}{;}{a}{}{,}
\usepackage{graphicx}
\date{} 
\title{\large\bf\flushleft The Statistical Analyses and The OPEA Model of The White-Light Flares Occurring on Kr\"{u}ger 60B (DO Cep)}
\author{\parbox{\textwidth}{\flushleft
\vspace{-0.5cm}
{\it Hasan Ali DAL\affil{*}}\\
\vspace{0.4cm}
{\small \affil{*}\,Department of Astronomy and Space Sciences, University of Ege, Bornova, 35100 ~\.{I}zmir, Turkey}\\
{\small \affil{*}\,Email: ali.dal@ege.edu.tr}}}
\begin{document}
\twocolumn[
\begin{changemargin}{.8cm}{.5cm}
\begin{minipage}{.9\textwidth}
\vspace{-1cm}
\maketitle
%
%
\small{\bf Abstract:}  In this study, new observations and some results of the statistical analyses are presented. The largest flare data set of DO Cep in the literature have been obtained with 89 flares detected in 67.61 hours $U$-band flare patrol. First of all, the observations demonstrated that the star is one of the most active flare stars in respect to the computed flare frequency. Secondly, using the independent samples t-test, the detected flares were classified into two subtypes, and then they were modelled. Analysing the models demonstrated that the fast and slow flares occurring on the star can be separated with a critical value of the ratio of their decay time to rise time. The critical value was computed as 3.40. According to this value, the fast flare rate is 20.22$\%$, while the slow flare rate is 79.78$\%$. Besides, there is a 39.282 times difference between the energies of these two types of flares. However, the flare-equivalent durations versus the flare rise times increase in similar ways for both groups. In addition, all the flares were modelled with the one-phase exponential association function. Analysing this model, the $plateau$ value was found to be 2.810. Moreover, the $half-life$ value was computed as 433.1 s from the model. The maximum flare rise time was found to be 1164 s, while the maximum flare total duration was found to be 3472 s. The results of the flare timescales indicate that the geometry of flaring loop on the surface of the star might be similar to those seen on analogues of DO Cep. Consequently, considering the both $half-life$ value and flare timescales, the flares detected on the surface of the DO Cep get maximum energy in longer times, while the geometries of flaring loops or areas get smaller.\\


\medskip
\medskip
\end{minipage}
\end{changemargin}
]
\small

\section{Introduction}

The white-light flares observed on the surfaces of UV Ceti-type stars and their process could not be absolutely understood, although these subjects have been heavily studied \citep{Ben10}. In this study, we obtained the largest data set from the observations of DO Cep in the literature. The data are very useful for a statistical analysis of the white-light flare property.

Observed star, DO Cep (= Kr\"{u}ger 60B = KR 60B), is classified as a UV Ceti-type star in the SIMBAD database from spectral-type dM4V \citep{Hen02}. The star is a component of HD 239960, which is a visual binary \citep{Lac77, Sod99}. The other component of the binary is GJ 860 A, which classified as dM3V by \citet{Hen02, Tam06}. Unlike DO Cep, GJ 860 A (= KR 60A) is not active as seen from the literature. In the studies of the system, the semi-major axis of the orbit ($a$) was found to be 2.420 $arcsec$, while the orbital inclination ($i$) was calculated as 172 $deg$ by \citet{Sod99}. The orbital period was found to be 44.64 years, while the orbit eccentricity ($e$) was computed as 0.41 in the same study. The distance of the system is given as 4.0 $pc$ by \citet{Pet91}, while it is given as 4.04 $pc$ by \citet{Sch04}. Some basic properties taken from \citep{Lac77} are listed in Table 1 for each component.

According to \citet{Vee74}, DO Cep is an old disk star. In fact, taking $M_{bol}=9.72$ mag and $log(T_{eff})=3.525$, its age was computed as about $5.0 \times 10^{8}$ years by \citet{Van83}. The equatorial rotational velocity ($v \sin i$) of DO Cep was found to be 4.7 $kms^{-1}$ by \citet{Gle05, Jen09}. There are several flare patrols for DO Cep in the literature \citep{Har55, Her69, Nic75, Con82}. For the first time, \citet{Har55} suspected that DO Cep is a flare star. Then, \citet{Her69} observed the star along 27.8 hours, and they detected 10 flares. Secondly, \citet{Nic75} detected 22 flares in 57.4 hours flare patrol, while \citet{Con82} detected no flare in 59.13 hours flare patrol. As seen from the literature, DO Cep has a high level flare activity.

In this study, DO Cep was observed in $U$-band for flare patrol in 2006 and 2007, and 89 white-light flares were detected. In order to classify the flares detected from DO Cep, the method described by \citet{Dal10} was used. In the literature, there are several studies about classifying the white-light flares \citep{Har69, Osa68, Mof74, Gur88}. The classification of the flare light-variations is important due to modelling these events \citep{Gur88, Ger05}. The white-light flare events were generally classified into two subtypes as slow and fast flares \citep{Har69, Osa68}. However, some studies, such as \citet{Osk69} and \citet{Mof74}, revealed that the white-light flares can be classified in more than two subtypes. \citet{Kun67} revealed that the observed flare light-curves should be a combination of slow and fast flares. Recently, \citet{Dal10} developed a rule to classifying white-light flares. The rule depended on the ratios of flare decay times to flare rise times demonstrates that the flare, whose decay time is 3.5 times longer than its rise time, is a fast flare. If the decay time of a flare is shorter than 3.5 times of its rise time, the flare can be classified as a slow flare. In fact, there are two possible energy sources for the white-light flares \citep{Gur88}. According to the author, the thermal processes are dominant in the slow flare events, while the nonthermal processes are dominant in the fast flare events. A rapid increasing is generally seen in light curves, if the energy source is caused by the nonthermal processes \citep{Ben10, Ger05, Gur88}.

Apart from the shapes of the flare light-variations, the upper and lower limits of both flare-power and flare timescales are also important to understand the flare processes occurring on a star. In order to compare the flare-powers of different stars, several studies have been done. In these studies, the flare energy spectra were derived for each star \citep{Ger72, Lac76, Wal81, Ger83, Pet84, Mav86}. According to the results of these studies, the energy levels of stars vary from a star to next one. The variations seem to be caused by different ages of stars. On the other hand, the analyses based on the flare energies could not give the real results. Because the flare energy depends on the luminosity of a star as well as the power of the flare. Thus, if the stars are from different spectral-types, the energies of these flare will be different, even if their real powers are the same. In this respect, the flare-equivalent duration was based on the analyses in this study in order to determined the behaviour of the white-light flares of DO Cep. These method recently developed by \citet{Dal11a} is based on the modelling the distributions of the flare-equivalent durations versus flare total durations. The authors demonstrated that the best function is the one-phase exponential association function (hereafter the OPEA) to model the distribution. As it is seen in the OPEA model, the flare-equivalent durations can not be higher than a specific value, and the flare's total duration does not matter. \citet{Dal11a} defined this level as an indicator for the saturation level for the white-light flare processes. In fact, white-light flares are detected in some large active regions, where compact and two-ribbon flares are occurring on the surface of the Sun \citep{Rod90, Ben10}. It is possibly expected that the energies or the flare-equivalent durations of white-light flares can also reach the saturation. Generally, flare activity seen on the surfaces of dMe stars is modelled concerning the processes of the solar flare event. This is why the magnetic reconnection process is accepted as the source of the energy in these events \citep{Ger05, Hud97}. According to both some models and observations, it is seen that some parameters of magnetic activity can reach the saturation \citep{Ger05, Sku86, Vil83, Vil86, Doy96a, Doy96b}.

\section{Observations and Analyses}

\subsection{Observations}

The observations of the flare patrol were acquired with a High-Speed Three Channel Photometer attached to the 48 cm Cassegrain type telescope at Ege University Observatory. Using a tracking star in second channel of the photometer, flare observations were continued in standard Johnson $U$-band with exposure times between 7 and 10 seconds in the time resolution of 0.01 seconds. Considering the technical properties of the HSTCP given by \citet{Mei02} and following the procedures outlined by \citet{Kir06}, the mean average of the standard deviations of observation times was computed as 0.08 seconds for the U band observations. Some properties of DO Cep and its comparisons are listed in Table 2. Standard $V$ magnitudes and $B-V$ color indexes obtained in this study are given in Table 2. Although DO Cep and its comparison stars are very close to one another on the celestial plane, differential extinction corrections were applied. The extinction coefficients were obtained from observations of the comparison stars on each night. Moreover, the variable and its comparison stars were observed in the standard Johnson $UBVR$ bands with the standard stars in their vicinity and the reduced differential magnitudes, in the sense of variable minus comparison stars, were transformed to the standard system using the procedures described by \citet{Har62}. The standard stars were chosen from the catalogues of \citet{Lan92}. Heliocentric corrections were applied to the times of the observations. The standard deviations of observation points acquired in the standard Johnson $UBVR$ bands are about 0.015 mag, 0.009 mag, 0.007 mag and 0.007 mag on each night, respectively. To compute the standard deviations of observations, we used the standard deviations of the reduced differential magnitudes in the sense comparisons (C1) minus check (C2) stars for each night. There is no variation in the standard brightness of the comparison stars. The flare patrol of DO Cep was continued for 9 nights between September 9 and November 26 in 2006, and 11 nights between July 31 and October 17 in 2007. The total duration of the $U$-band flare patrol is 22.76 h in 2006, while it is 44.85 h in 2007. 4446 observing pints were obtained in 2006, while 9841 observing points were obtained in 2007. According to the $3\sigma$ of the $U$ band standard deviation in each night, it was decided whether an event observed in that night is a flare, or not. Therefore, 88 flares were detected in 2007, while only one flare could be detected in 2006.

\citet{Ger72} developed a method for calculating flare energies. Flare-equivalent durations (s) and energies (erg) were calculated using equations (1) and (2) of this method,

\begin{center}
\begin{equation}
P = \int[(I_{flare}-I_{0})/I_{0}] dt
\end{equation}
\end{center}
where $I_{0}$ is the intensity of the star in the quiescent level and $I_{flare}$ is the intensity during flare, and

\begin{center}
\begin{equation}
E = P \times L
\end{equation}
\end{center}
where $E$ is the flare energy (erg), $P$ is the flare-equivalent duration (s), and $L$ is the luminosity of the stars in the quiescent level in the Johnson $U$-band. It must be noted that the flare-equivalent duration has a time unit (s).

The parameters of the flare light curves were calculated for each flare. All the parameters were computed following the procedure described in detail by \citet{Dal10}. There are some important points in the procedure. We firstly separated each flare light curve into three parts. One of them is the part indicating the quiescent level of the brightness before the first flare on each nigh. The brightness level without any variations (such as a flare or any oscillation) was taken as a quiescent level of the brightness of this star. To determine this level, we used the standard deviation of each observation point, considering the mean average of all the observation points until this last point. If the standard deviations of that one and following points get over the $3\sigma$ level, this point was taken as the beginning of a flare. Thus, the quiescent levels of each star were determined from all the observation points before the first flare on each night. We fitted this level with a linear function, and then, using this linear function, we computed the flare-equivalent duration, flare amplitude and all the flare time-scales (rise and decay times). The part of the light curve above the quiescent level was also separated into two sub-parts. First of them is the impulsive phase, in which the flare increases. Second one is the decay phase. The impulsive and decay phases were separated according to the maximum brightness observed in this part. It must be noted that some flares have a few peaks. In this case, the point of the first-highest peak was assumed as the flare maximum. To determine the flare time-scales, we fitted the impulsive and decay phases with the polynomial functions. The best polynomial functions were chosen according to the correlation coefficients ($r^{2}$) of fits. To determine the beginning and end of each flare, we computed the intersection points of the polynomial fits with the linear fit of the quiescent level and their standard deviations. In this study, the intersection points were taken as the beginning and end of each flare. The flare rise time was taken the duration between the beginning and the flare maximum point. In the same way, the flare decay time was taken the duration between the flare maximum point and the flare end. The height of the observed-maximum points from the quiescent level was taken as the amplitudes of this flare. The same procedure was used for each flare, and Grahp-Pad Prism V5.02 \citep{Mot07} software was performed in all calculations.

All calculated parameters are listed in Table 3 for 89 flares. The observing date, HJD of flare maximum time, flare rise and decay times (s), flare total duration (s), flare-equivalent durations (s), flare amplitude in $U$-band (mag), $U-B$ color index (mag), flare energy (erg) and flare type are listed in the columns of the table, respectively. As it is explained in Section 1, other important point is that the flare-equivalent durations were used in the analyses due to the luminosity term ($L$) in equation (2), in stead of flare energies.

\begin{center}
\begin{equation}
N~=~\Sigma n_{f}~/~\Sigma T_{t}
\end{equation}
\end{center}

In addition to these parameters, following the method used by \citet{Let97} and \citet{Dal11b}, the flare frequency ($N$) was computed for each observing season. In equation (3), $n_{f}$ is the total number of the flare detected in a season, and $T_{t}$ is total time of the flare patrol in that season. Using equation (3), the value of $N$ was found to be 0.044 for 2006. It was found to be 3.866 for 2007.

If the detected flares are examined, it will be seen that the light curve of each flare has a distinctive light-variation shape. Five light curve parts from the observations are seen in Figures 1 - 5 for the examples. The horizontal dashed lines seen in these figures represent the level of quiescent brightness. Three flares detected on 2007, August 1 are seen in Figure 1. According to the rule described by \citet{Dal10}, the Flare A and B shown in Figure 1 are two slow flare samples, while the Flare C is a sample of the fast flare. The flares seen in Figure 2 were detected on 2007, August 3. In this figure, the Flare A and B are the samples of the fast flares, while the Flare C is a slow flare. Figure 3 shows two flares, and both of them are the fast flares. The Flare A in this figure is the most powerful flare detected in this study. Its amplitude is 1.90 mag in the U band. Its rise time is 270 s, and decay time is 3202 s. The flare seen in figure 4 is a combined flare. There should be actually two flares (Part A and Part B), but their light variations are combined in the light curves. \citet{Mof74} classified the flares like this as a complex flares. It should be noted that the flares like this one were not ignored in the analyses described in Section 2.2. Another interesting samples are seen in Figure 5. In the figure, Flare A and B are the slow flare, while Flare C is the fast flare. Apart from these three flares, there are three spikes, which are combined with Flare A and B.

\subsection{Fast and Slow Flares}

The flares detected from DO Cep were analysed using the method developed by \citet{Dal10}. It was tested whether the limit ratio (3.50) is also acceptable for the white-light flares detected from DO Cep. Thus, using new-large data set, it was tested whether the value of 3.50 is a general limit, or not.

In first step, the equivalent durations of flares, whose rise times are equal, were compared. For example, there are 13 flares, whose rise time are 15 s. The light variations of 9 flares among these 13 flares are similar to Flare A and B seen in Figure 1. The light variations of other 4 fares among these 13 flares are similar to Flare C shown in Figure 1. The average of their equivalent durations is 6.206 s for 9 flares, which are similar Flare A and B. However, the average of their equivalent durations is 29.551 s for other 4 flares. The main difference of these two example groups is seen in the shapes of the light curves. Finally, we found 18 flares with higher energy and 71 flares with lower energy among 83 flares detected from DO Cep.

Using the independent samples t-test (hereafter t-test) \citep{Wal03, Daw04} in the SPSS V17.0 \citep{Gre99} and Grahp-Pad Prism V5.02 \citep{Mot07} software, data sets were analysed in order to test whether these two groups are statistically independent from each other. In the analyses, the flare rise times were taken as a dependent variable, and the flare-equivalent durations were taken as an independent variable. The value of ($\alpha$) is taken as 0.005, which allowed us to test whether the results are statistically acceptable, or not \citep{Daw04}.

The mean average of the equivalent durations for 71 slow flares was found to be 1.544$\pm$0.067 s, and it was computed as 1.871$\pm$0.130 s for 18 fast flares in the logarithmic scale. This shows that there is a difference of about 0.327 s between average equivalent durations in the logarithmic scale. The probability value (hereafter $p-value$) was computed to test the results of the t-test, and it was found to be $p<0.0001$. Considering $\alpha$ value, this means that the result is statistically acceptable. All the results obtained from the t-test analyses are given in Table 4.

In the second step, the distributions of the equivalent durations ($logP_{u}$) versus flare rise times ($logT_{r}$) were modelled for both flare types. Using the least-squares method, the best models for the distributions were examined in the SPSS V17.0 and Grahp-Pad Prism V5.02 software. The regression calculations demonstrated that the best fits of distributions are linear functions. The derived linear fits given by equations (4) and (5) are shown in Figure 6.

\begin{center}
\begin{equation}
log(P_{u})~=~1.232~\times~log(T_{r})~-~0.137
\end{equation}
\end{center}

\begin{center}
\begin{equation}
log(P_{u})~=~1.046~\times~log(T_{r})~-~0.450
\end{equation}
\end{center}

In the next step, the linear functions were compared. The slope of the linear function was found to be 1.046$\pm$0.048 for slow flares, while it was computed as 1.232$\pm$0.181 for fast flares. Then, the $p-value$ was calculated and found to be $p=0.650$. The $p-value$ indicates that there is no significant difference between the slopes of fits, and it can be assumed that they are statistically parallel.

Finally, the $y-intercept$ values of both linear fits were compared. The $y-intercept$ value was found to be -0.450 for the slow flares, and it was found to be -0.137 for the fast flares in the logarithmic scale. There is a difference of about 0.313 between them. Then, the $p-value$ was computed for the $y-intercept$ values to say whether there is a statistically significant difference, it was found to be $p<0.0001$. The result demonstrated that the difference between two $y-intercept$ values is obviously important.

The distributions of the equivalent durations in the logarithmic scale versus flare rise times were modelled with the OPEA function to find the maximum energy levels and timescales of the two flare types. The derived distributions are shown in Figure 7. Using the least-squares method, the regression calculations showed that the averaged value of upper limit is 2.559$\pm$0.096 for the slow flares. On the other hand, the regression calculations indicated that the distribution can not be modelled with the OPEA function due to the linear increasing. The linear fit derived for the fast flares is also seen in Figure 7. Beside the averaged value of upper limit, it was found that the lengths of rise times for slow flares can reach to 1164 s, while they are not longer than 270 s for fast flares.

\subsection{The One-Phase Exponential Association Models of the Distribution of the Flares}

The distribution of the flare-equivalent durations in the logarithmic scale versus the flare total durations indicates that the flare mechanism occurring on the surface of DO Cep has a upper limit for the flare energy. To examine this case, the distribution was modelled and statistically analysed. First of all, the distribution of the equivalent durations ($logP_{u}$) in the logarithmic scale versus the flare total durations was obtained. Then, using regression calculations, the best function was determined to fit the distribution by SPSS V17.0 software. The regression analyses demonstrated that the OPEA function \citep{Mot07, Spa87} given by equation (6) is the best model fit. According to \citet{Dal11a}, this is actually an expected case. The case demonstrates that the flares occurring on the surface of DO Cep have an upper limit for producing energy. In the final step, the OPEA model of the distributions was derived by Grahp-Pad Prism V5.02 using the least-squares method.

\begin{center}
\begin{equation}
y~=~y_{0}~+~(plateau~-~y_{0})~\times~(1~-~e^{-k~\times~x})
\end{equation}
\end{center}

The details of the OPEA function have been given by \citet{Dal11a}. In brief, some important parameters can be derived from the OPEA function, and these parameters reveal the condition of the flare mechanism occurring on the surface of the star. One of them is $y_{0}$, which is the lower limit of equivalent durations for observed flares in the logarithmic scale. In contrast to $y_{0}$, the parameter of $plateau$ is the upper limit. It should be noted that the $y_{0}$ parameter depends on the quality of observations as well as flare power. However, $plateau$ parameter depends only on power of flares. \citet{Dal11a} identified the $plateau$ parameter as a saturation level for the white-light flare activity observed in $U$-band. The derived OPEA model is shown in Figure 8, while the parameters of the model are listed in Table 5. The $span$ value listed in the table is the difference between the values of $plateau$ and $y_{0}$. One of the most important parameters is the $half-life$ value. This parameter is half of the first $x$ values, where the model reaches the $plateau$ values. In other words, it is half of the flare total duration, where flares with the highest energy start to be seen. In order to test the the $plateau$ values derived from the OPEA model, the upper limit of the equivalent durations was computed using the t-test. Thus, the $plateau$ value was tested whether it is statistically acceptable, or not. The flares in the $plateau$ phases of the model were only used to test. The mean average of the equivalent durations was computed and found to be 2.808$\pm$1.149.

As seen from the data distributions, the maximum flare rise time obtained from these 89 flares is 1164 s, while the maximum flare total duration is 3472 s.

\section{Results and Discussion}

\subsection{Flare Activity and Flare Types}

In this study, 89 white-light flares were detected in $U$-band observations of DO Cep. 88 flare were detected in 44.85 h flare patrol of 2007, while only one flare was detected in 22.76 h flare patrol of 2006. Therefore, 0.044 flares were detected per hour in 2006, while 3.866 flares were detected per hour in 2007. There is a large difference between the flare frequencies ($N_{2006}$ and $N_{2007}$) of consecutive observing seasons. A large differences between the flare frequencies obtained in different years are also seen in the literature \citep{Her69, Nic75, Con82}. According to the results of \citet{Her69}, $N$ value is 0.360 in 1968. The flare frequency ($N$) is 0.509 flares per hour in 1970, and it is 0.208 flares per hour in the observing season of 1972 - 1973 \citep{Nic75}. However, \citet{Con82} detected no flare in 1975. Consequently, the largest frequency was obtained with 3.866 flares per hour in this study. It seems that DO Cep is active as well as EV Lac, EQ Peg and AD Leo \citep{Mof74, Dal10}. The flare frequency variation of UV Ceti type stars has been examined in several studies \citep{Ish91, Let97}. \citet{Ish91} found no variation for a few stars, while \citet{Let97} demonstrated that the flare frequency of EV Lac is dramatically increasing. It must be noted that DO Cep should be taken to observing programs, because its flare frequency is remarkably varying.

The flares detected in this study are examined one by one. The flares, whose rise times are equal, were determined. It was seen that the flares are accumulating into two groups. It was seen that even if the rise times of two flares are equal, their equivalent durations can be different from each other. Apart from their equivalent durations, the main difference between two type flares is light-variation shapes. As seen from Figures 1 - 5, the some flares slowly increase and slowly decrease, while some of them rapidly increase, but slowly decrease.

In logarithmic scale, the flare distributions were obtained for both groups. First of all, two group flares were analysed with t-test. Then, they were modelled with the linear function, and the models of two groups were analysed to compare. Using t-test, the averages of equivalent durations for two types of flares were computed. The average of equivalent durations was found to be 1.871 s for the fast flares, and it is 1.544 s for the slow flares. The difference of 0.327 s between these values in the logarithmic scale is equal to the 39.282 s difference between the equivalent durations. As can be seen from equation (2), this difference between average equivalent durations affects the energies in the same way. Therefore, there is 39.282 times difference between the energies of these two types of flares. This difference must be the difference mentioned by \citet{Gur88}. On the other hand, according to \citet{Dal10}, this difference between two flare types is about 157 times. In the case of DO Cep, the energies of the slow and fast flares occurring on the surface of the star seem to be closer to each other.

Apart from the average of equivalent durations, the parameters of the linear fits were also compared. The slope of the linear fit is 1.046 for the slow flares, which are low-energy flares, and it is 1.232 for the fast flares, which are high-energy flares. According to the $p-value$, the slopes are almost close to each other. It demonstrates that the flare-equivalent durations versus the flare rise times increase in similar ways for both groups. However, the fast flare (Flare A) seen in Figure 3 is seen out of the general trend of the fast flares. This flare is the most powerful flare detected in the study. It seems to be an extreme example. In the case of the extreme examples, some effects must be involved in the fast flare process towards the long rise times. These effects can make fast flares seem more powerful than they actually are. However, comparing the $y-intercept$ values of the linear fits, it is seen that there is a 0.313 times difference in the logarithmic scale. There is a 0.327 times difference between general averages. Both values are close to each other. It means that the energy emitting processes behave similarly except the extreme flare. Apart from the equivalent durations, the differences between these two flare types are seen in the lengths of their rise times and their amplitudes. The maximum rise time seen among the slow flares is 1164 s, but it is 270 s for fast flares. In addition, the amplitudes of slow flares can reach to 0.791 mag at most, the amplitudes of fast flares can reach to 1.900 mag.

\citet{Dal10} computed the ratios of flare decay times to flare rise times for two types of flares. They have demonstrated that there is a limit values between two flare types. This limit value of the ratio of flare decay time to flare rise time is 3.50. In this study, the limit value of this ratio is found to be 3.40 for the flares detected from DO Cep. Providing this limit value between flare types, it was found that the fast flare rate is 20.22$\%$ of the 89 flares observed in this study, while the slow flare rate is 79.78$\%$. It means that one of every five flares is the fast flare, other four flares are the slow flare. This result is close to what \citet{Gur88} stated. According to \citet{Gur88}, slow flares with low energies and low amplitudes make up 95$\%$ of all flares, and the remainder are fast flares.

\subsection{The Saturation Level in the Detected White-Light Flare}

The distributions of flare-equivalent durations versus flare total duration were modelled by the OPEA function expressed by equation (6) for 89 white-light flares detected in observations of DO Cep. To model the distribution, the best model curve was searched. Considering $p-value$ and the correlation coefficient ($r^{2}$) parameters, the OPEA function was found as the best model function. The main characteristic feature of the OPEA is that this function has a $plateau$ phase. According to the observations, the flare-equivalent durations increase with the flare total duration until a specific total duration value. After the specific total duration, the flare-equivalent durations are constant, and the total duration does not matter. There is just one flare among all of them. This flare is Flare A seen in Figure 3. It must be an extreme sample.

Some parameters such as $plateau$ value, $half-life$, etc., were derived from the OPEA model. The $plateau$ value was found to be 2.810. The value is in agreement with the mean average of the equivalent durations. It had been found to be 2.808. Considering the standard deviations of two values, they can be assumed to be equal. Besides, the found $plateau$ value is also in agreement with the $plateau$ values found from other stars by \citet{Dal11a}. According to the $B-V$ colour index of DO Cep, it is seen that the star is among its analogues. This result supports that the upper limit of the energy producing by white-light flare mechanism really increase towards the later spectral types.

It is well known that the white-light flares occur in the regions, where the compact and two-ribbon flare events are seen \citep{Rod90, Ben10}. In the analyses, the flare-equivalent durations were used instead of the flare energies. In fact, the derived $plateau$ values depend only on the power of the white-light flares. According to observations, the $plateau$ phase exists in the model. The flare-equivalent durations can not be higher than a particular value, and the flare's total duration does not matter. Apart from the timescales, the power of the flares must depend on some other parameters, such as magnetic field flux and/or particle density in the volumes of the flare processes. However, \citet{Doy96a} and \citet{Doy96b} suggested that the saturation in the active stars does not have to be related to the filling factor of magnetic structures on the stellar surfaces or the dynamo mechanism under the surface. It can be related to some radiative losses in the chromosphere, where the temperature and density are increasing in the case of fast rotation. This phenomenon can occur in the chromosphere due to the flare process instead of fast rotation, and this causes the $plateau$ phase to occur in the distributions of flare-equivalent duration versus flare total duration. On the other hand, the $plateau$ phase cannot be due to some radiative losses in the chromosphere with increasing temperature and density. This is because \citet{Gri83} demonstrated the effects of radiative losses in the chromosphere on the white-light photometry of the flares. According to \citet{Gri83}, the negative H opacity in the chromosphere causes the radiative losses, and these are seen as pre-flare dip in the light curves of the white-light flares. Unfortunately, considering the results of \citet{Dal11a}, it is seen that the $plateau$ values vary from one star to the next. This indicates that some parameters or their efficacies, which make the $plateau$ increase, are changing from star to star. According to Standard Magnetic Reconnection Model developed by \citet{Pet64}, there are several important parameters giving shape to flare events, such as Alfv\'{e}n velocity ($\nu_{A}$), $B$, the emissivity of the plasma ($R$) and the most important one, the electron density of the plasma ($n_{e}$) \citep{VanB88, VanA88}. All these parameters are related with both heating and cooling processes in a flare event. \citet{VanA88, VanB88} have defined the radiative loss timescale ($\tau_{d}$) as $E_{th}/R$. Here $E_{th}$ is the total thermal energy, while $R$ is emissivity of the plasma. $E_{th}$ depends on the magnetic energy, which is defined as $B^{2}/8\pi$, and $R$ depends on the electron density ($n_{e}$) of the plasma. $\tau_{d}$ is firmly correlated with $B$ and $n_{e}$, while $\tau_{r}$ is proportional to a larger loop length ($\ell$) and smaller $B$ values. Consequently, it is seen that both the shape and power of a flare event depend on mainly two parameters, $n_{e}$ and $B$.

As seen from the OPEA model, the flare-equivalent durations start to reach maximum value in a specific total duration, and the $half-life$ value was found to be 433.1 s from the model. In addition, the maximum flare rise time was found to be 1164 s, while the maximum flare total duration was found to be 3472 s. These results demonstrated that the flare timescales of the flares detected from DO Cep are in fact shorter than they are in the earlier spectral types. However, the flares get the maximum energy limits in longer times. It is well known from the X-Ray observations of the flares that the timescales of the X-Ray flares give some clues about the flaring loop geometry on the stars \citep{Ree02, Ima03, Fav05, Pan08}. The white-light flares can exhibit the same behaviour with its counterpart observed in X-Ray \citep{Ger05, Ben10}. If this case is valid, the timescales derived from the white-light flares can also give some clues about the the flaring loop geometry or the flaring area geometry (at least for the photosphere). The obtained timescales from the observations of DO Cep demonstrated that the flaring loop or area are smaller than those seen on the stars from the earlier spectral types. Because, the obtained maximum flare duration for DO Cep flares is 3472 s. The observed maximum duration is 5236 s for V1005 Ori, and 4164 s for AD Leo \citep{Dal11a}. The flare timescales of both stars are dramatically longer than that of DO Cep. However, considering the $half-life$ value, the flares detected from DO Cep get maximum energy in longer times, while the geometries of flaring loops or areas get smaller.

\section*{Acknowledgments} The author acknowledges generous allotments of observing time at the Ege University Observatory. I also thank the referee for useful comments that have contributed to the improvement of the paper.

\clearpage

\begin{figure*}[h]
\begin{center}
\includegraphics[scale=0.75, angle=0]{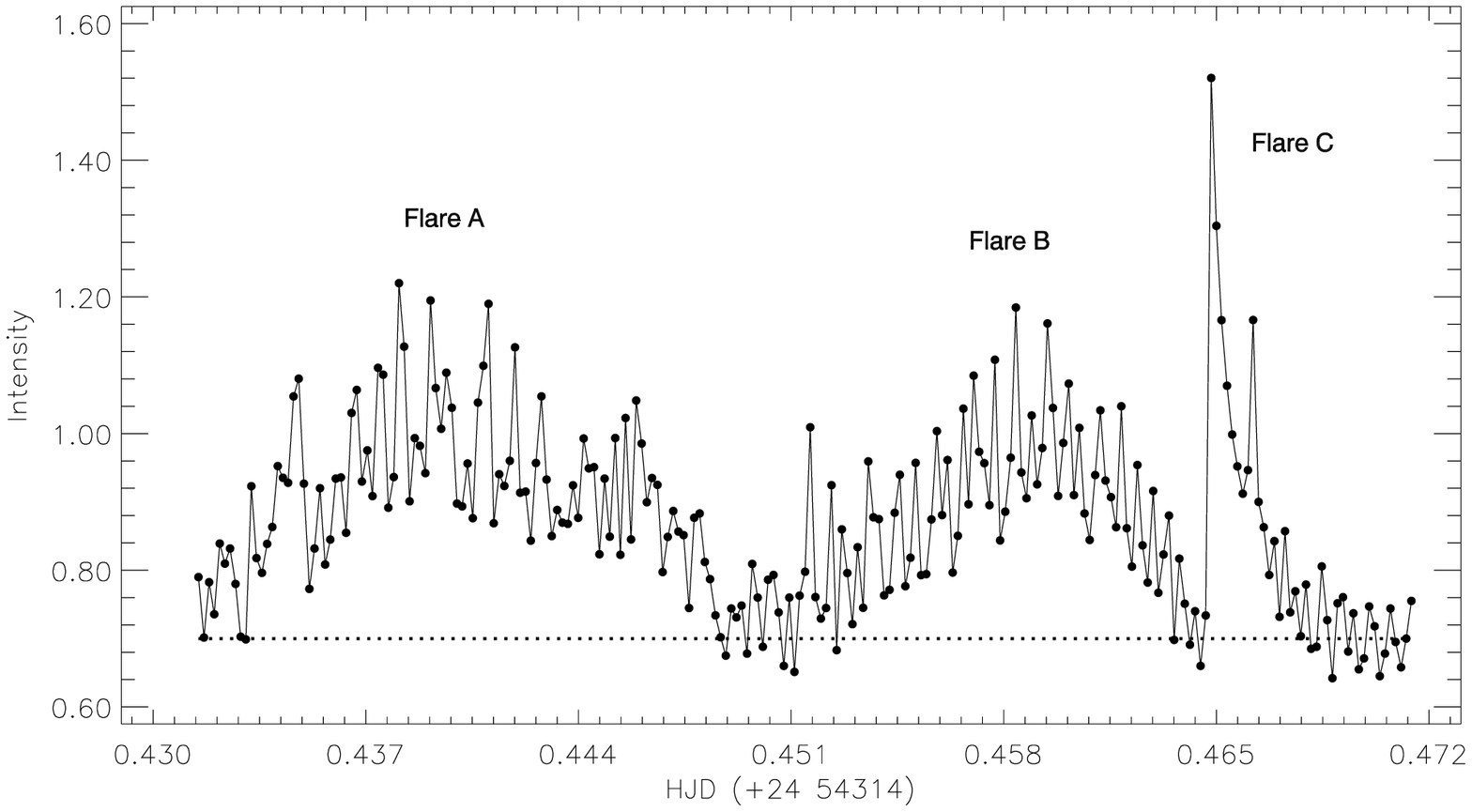}
\vspace{0.3cm}
\caption{The flare light curves detected in U-band observations of DO Cep on 2007, August 1. In figure, Flare A and B are two samples for the slow flares, while Flare C is a sample for the fast flares.}
\label{Fig1}
\end{center}
\end{figure*}

\begin{figure*}[h]
\begin{center}
\includegraphics[scale=0.75, angle=0]{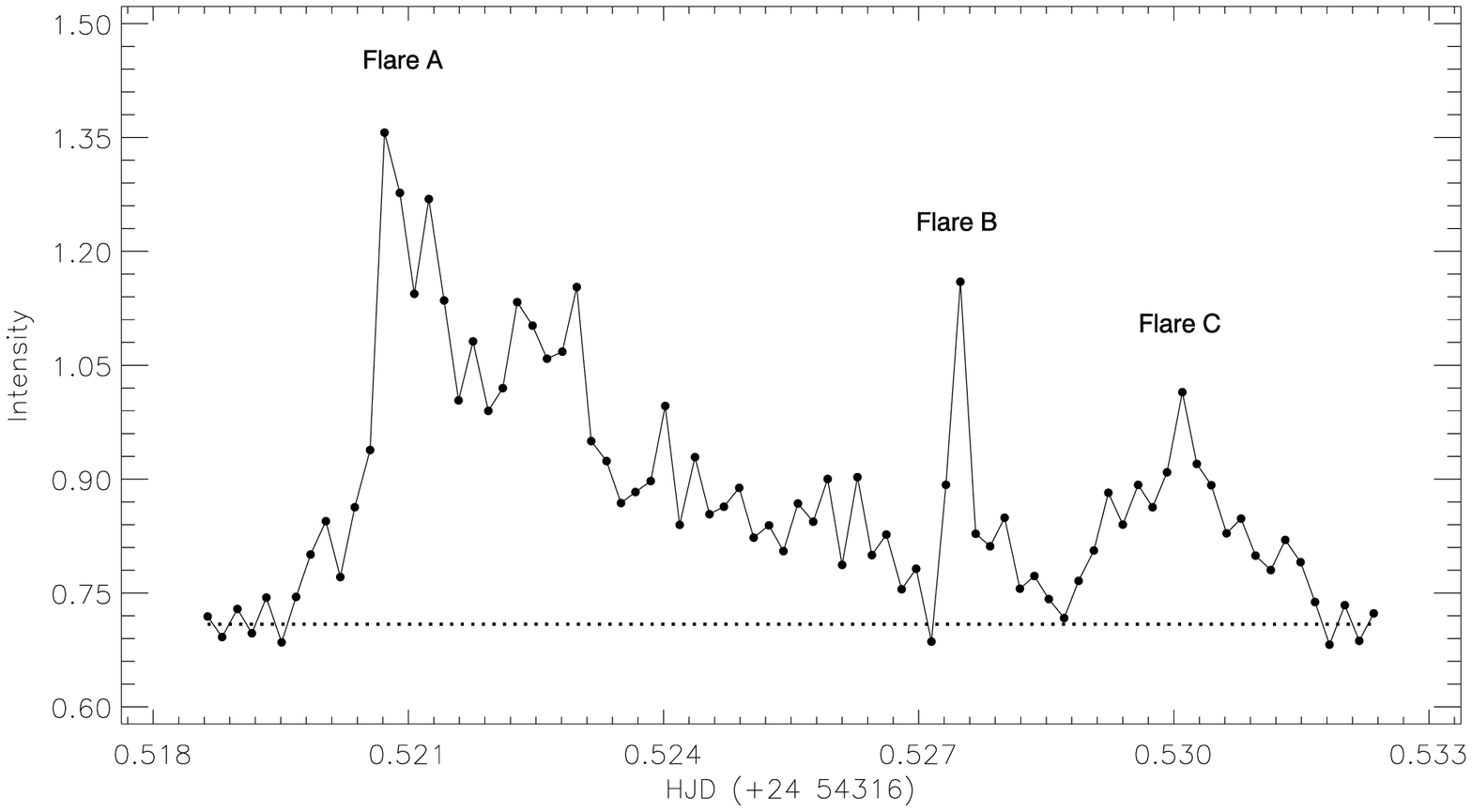}
\vspace{0.3cm}
\caption{The flare light curves detected in U-band observations of DO Cep on 2007, August 3. In figure, Flare A and B are two samples for the fast flares, while Flare C is a sample for the slow flares.}
\label{Fig2}
\end{center}
\end{figure*}

\begin{figure*}[h]
\begin{center}
\includegraphics[scale=0.75, angle=0]{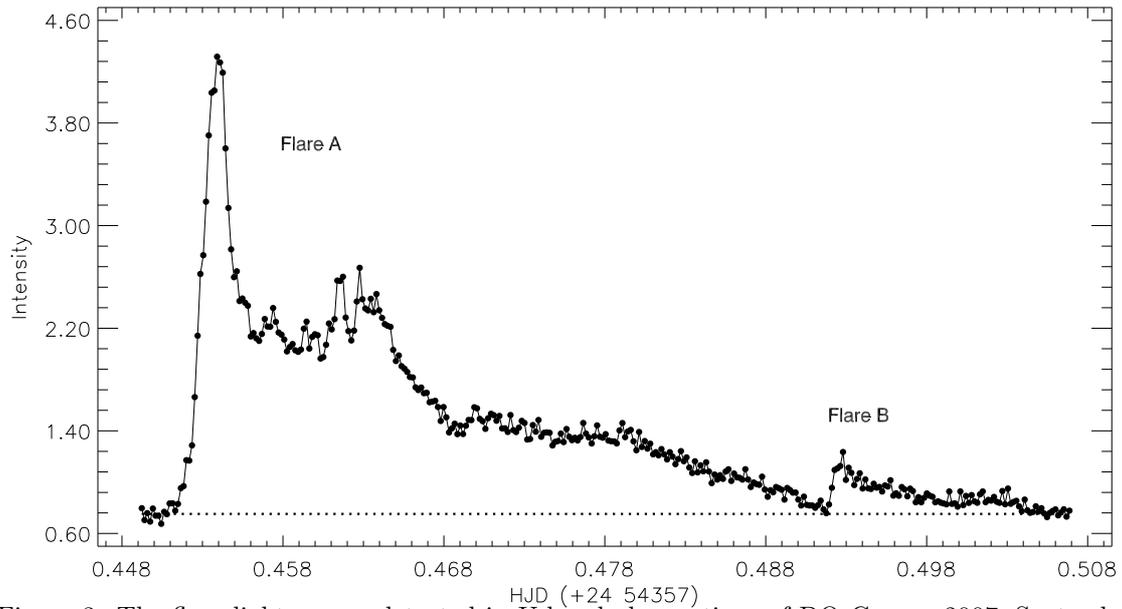}
\vspace{0.3cm}
\caption{The flare light curves detected in U-band observations of DO Cep on 2007, September 13. In figure, both Flare A and B are two samples for the fast flares.}
\label{Fig3}
\end{center}
\end{figure*}

\begin{figure*}[h]
\begin{center}
\includegraphics[scale=0.75, angle=0]{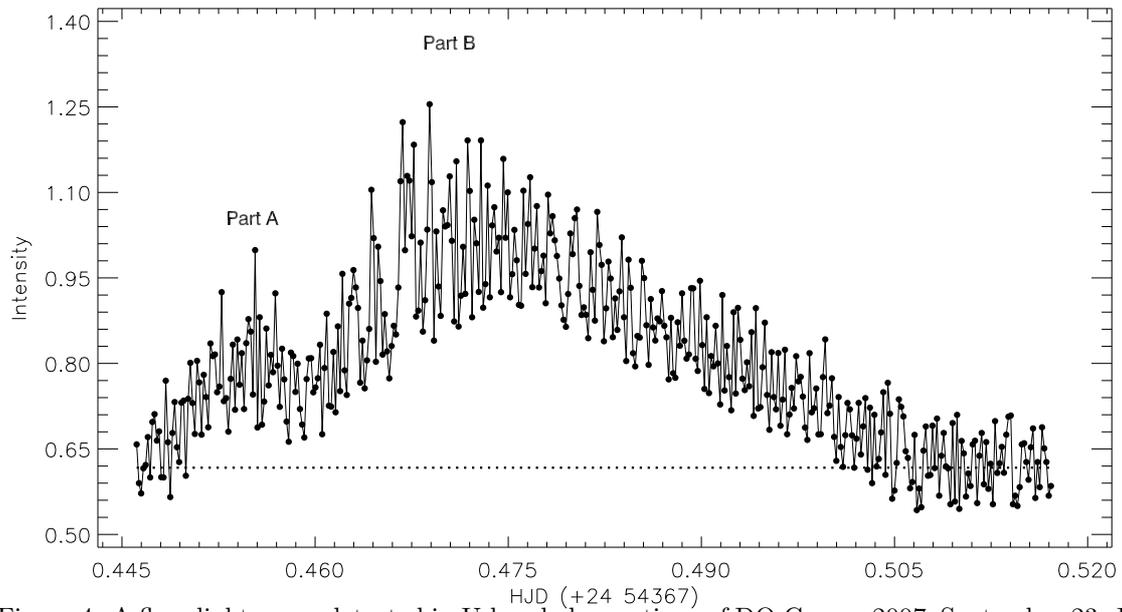}
\vspace{0.3cm}
\caption{A flare light curve detected in U-band observations of DO Cep on 2007, September 23. In figure, this flare light curve is a sample for the combined flares. Part A is a flare part, while Part B is a part of another flare. Both flares would be probably a slow flare.}
\label{Fig4}
\end{center}
\end{figure*}

\begin{figure*}[h]
\begin{center}
\includegraphics[scale=0.75, angle=0]{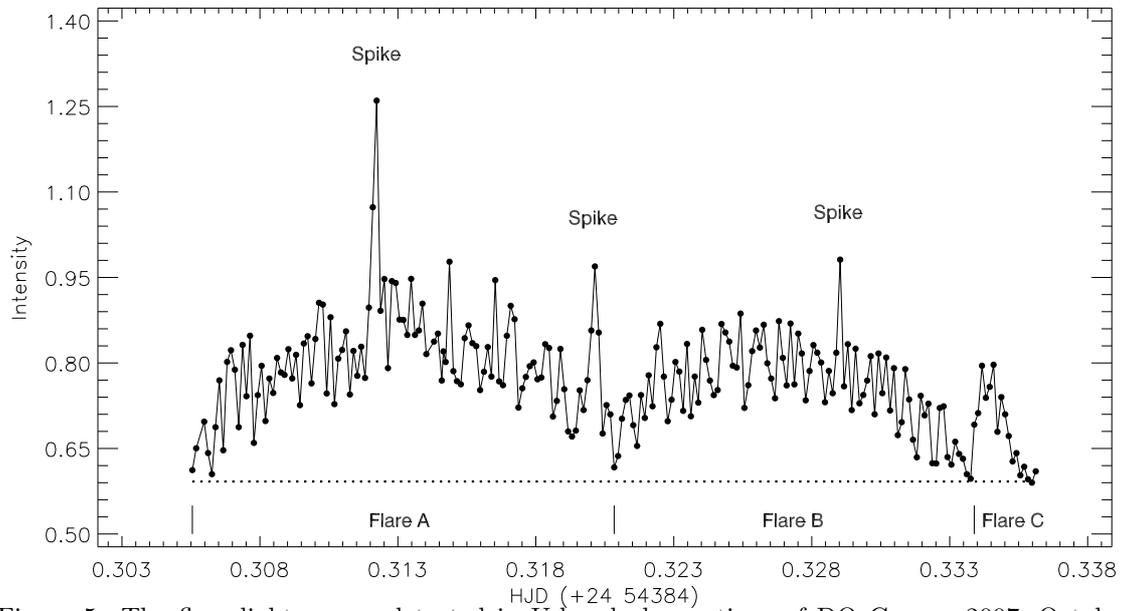}
\vspace{0.3cm}
\caption{The flare light curves detected in U-band observations of DO Cep on 2007, October 10. In figure, all three flares are the samples for the slow flares. Besides, three impulsive spikes are seen during the first two slow flares.}
\label{Fig5}
\end{center}
\end{figure*}

\begin{figure*}[h]
\begin{center}
\includegraphics[scale=0.75, angle=0]{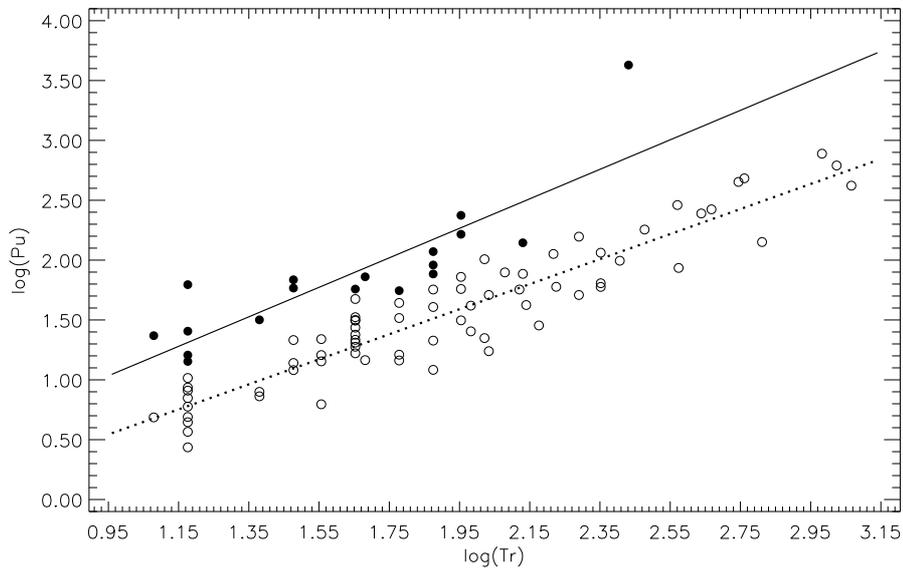}
\caption{Distributions for the mean averages of the equivalent durations ($logP_{u}$) vs. flare rise times ($logT_{r}$) in the logarithmic scale. In the figure, open circles represent slow flares, while filled circles show the fast flares. Lines represent fits given in equations (4) and (5).}
\label{Fig6}
\end{center}
\end{figure*}

\begin{figure*}[h]
\begin{center}
\includegraphics[scale=0.75, angle=0]{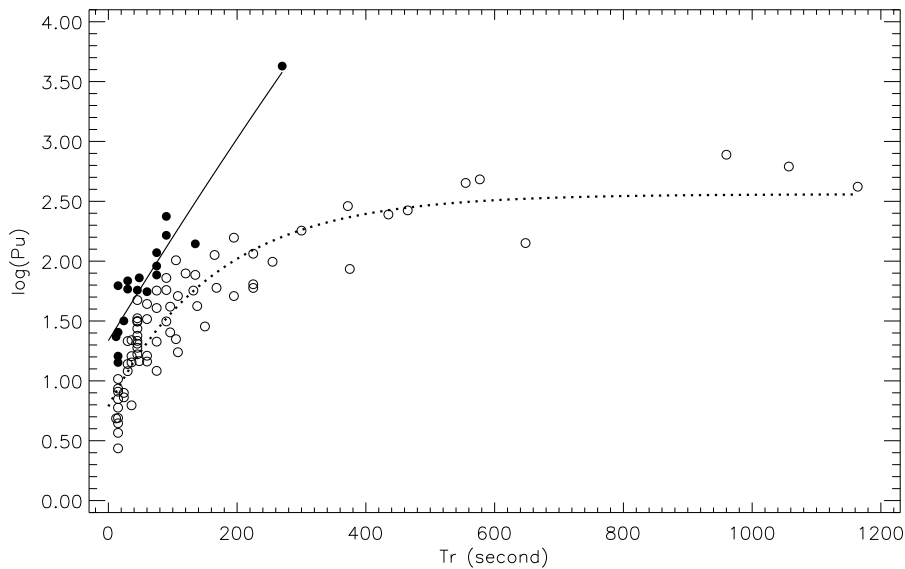}
\caption{Distributions of the equivalent durations ($logP_{u}$) in the logarithmic scale vs. flare rise times ($T_{r}$) for all 89 flares detected in observations of program stars. In the figure open circles represent slow flares, while filled circles show the fast flares.}
\label{Fig7}
\end{center}
\end{figure*}

\begin{figure*}[h]
\begin{center}
\includegraphics[scale=0.75, angle=0]{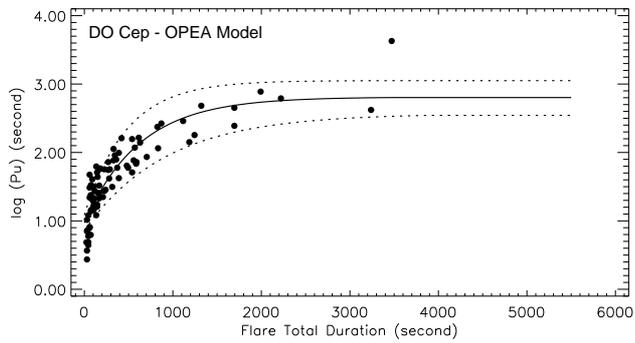}
\caption{Distributions of flare-equivalent duration on a logarithmic scale vs. flare total duration. Filled circles represent equivalent durations computed from flares detected from DO Cep. The line represents the model identified with equation (6) computed using the least-squares method. The dotted lines represent 95$\%$ confidence intervals for the model.}
\label{Fig8}
\end{center}
\end{figure*}

\clearpage

\begin{table*}[h]
\begin{center}
\caption{Basic parameters for the components of visual binary Kr\"{u}ger 60. All the parameters were taken from \citet{Lac77}.\label{tbl-1}}
\begin{tabular}{@{}lcc@{}}
\hline\hline
\textbf{Parameter}	&	\textbf{KR 60A} &	\textbf{KR 60B}	\\
	& (GJ 860 A) & (DO Cep)	\\
\hline					
V (mag)	&	9.850	&	11.220	\\
V-R (mag)	&	1.760	&	1.890	\\
$\pi \times 10^{3}$ (arcsec)	&	251$\pm$5	&	251$\pm$5	\\
$log(R_{star}/R_{\odot})$	&	-0.45$\pm$0.05	&	-0.65$\pm$0.05	\\
$log(M_{star}/M_{\odot})$	&	-0.57$\pm$0.03	&	0.80$\pm$0.03	\\
\hline
\end{tabular}
\end{center}
\end{table*}

\begin{table*}
\begin{center}
\caption{Basic parameters for the star studied and its comparison (C1) and check (C2) stars.\label{tbl-2}}
\begin{tabular}{@{}lcc@{}}
\hline\hline
\textbf{Stars}	&	\textbf{V (mag)}	&	\textbf{B-V (mag)}	\\
\hline					
DO Cep	&	9.615	&	1.604	\\
C1 = HD 239952	&	9.528	&	1.378	\\
C2 = SAO 34476	&	7.943	&	0.530	\\
\hline
\end{tabular}
\end{center}
\end{table*}

\setcounter{table}{2}
\begin{table*}
\setlength{\tabcolsep}{0.3 pt}
\centering
\caption{The parameters derived from analyses of the detected flares.\label{tbl-3}}
\begin{tabular}{@{}cccccccccc@{}}
\hline\hline
Observing	&	HJD of Flare & Rise &	Decay &	Total &	Equivalent	&	Flare	&	Flare	&	Flare	&	Flare	\\
 &	Maximum	& Time	&	Time &	Duration & Duration	& Amplitude	& U-B	& Energy	&		\\
Date	&	(+24 00000)	&	(s)	&	(s)	&	(s)	&	(s)	&	(mag)	&	(mag)	&	(erg)	&	Type  \\
\hline
09.10.2006	&	 54049.31335     	&	45	&	75	&	120	&	27.31375	&	0.386	&	0.933	&	1.28775E+31	&	Slow	\\
31.07.2007	&	 54313.46502     	&	45	&	30	&	75	&	23.82013	&	0.725	&	0.831	&	1.12303E+31	&	Slow	\\
31.07.2007	&	 54313.50060     	&	45	&	15	&	60	&	47.32471	&	0.609	&	0.919	&	2.23119E+31	&	Slow	\\
31.07.2007	&	 54313.50355     	&	75	&	15	&	90	&	40.59164	&	0.530	&	0.909	&	1.91375E+31	&	Slow	\\
31.07.2007	&	 54313.51657     	&	960	&	1032	&	1992	&	774.43561	&	0.538	&	0.715	&	3.65119E+32	&	Slow	\\
01.08.2007	&	 54314.43809     	&	555	&	1140	&	1695	&	449.94280	&	0.601	&	0.867	&	2.12132E+32	&	Slow	\\
01.08.2007	&	 54314.46118     	&	465	&	405	&	870	&	265.55134	&	0.539	&	0.832	&	1.25198E+32	&	Slow	\\
01.08.2007	&	 54314.46604     	&	15	&	120	&	135	&	62.38419	&	0.487	&	0.851	&	2.94119E+31	&	Fast	\\
01.08.2007	&	 54314.46760     	&	15	&	45	&	60	&	30.62136	&	0.791	&	0.596	&	1.44369E+31	&	Slow	\\
03.08.2007	&	 54316.44609     	&	300	&	945	&	1245	&	179.91426	&	0.261	&	1.075	&	8.48232E+31	&	Slow	\\
03.08.2007	&	 54316.46772     	&	15	&	30	&	45	&	4.42329	&	0.230	&	1.135	&	2.08543E+30	&	Slow	\\
03.08.2007	&	 54316.48127     	&	15	&	15	&	30	&	7.04602	&	0.331	&	1.008	&	3.32195E+30	&	Slow	\\
03.08.2007	&	 54316.48213     	&	45	&	60	&	105	&	18.89710	&	0.280	&	1.094	&	8.90931E+30	&	Slow	\\
03.08.2007	&	 54316.48682     	&	60	&	60	&	120	&	16.21439	&	0.284	&	1.089	&	7.64451E+30	&	Slow	\\
03.08.2007	&	 54316.50990     	&	45	&	60	&	105	&	16.65787	&	0.295	&	1.008	&	7.85360E+30	&	Slow	\\
03.08.2007	&	 54316.51355     	&	45	&	60	&	105	&	21.59144	&	0.055	&	1.007	&	1.01796E+31	&	Slow	\\
03.08.2007	&	 54316.52072     	&	90	&	735	&	825	&	236.56860	&	0.651	&	0.711	&	1.11534E+32	&	Fast	\\
03.08.2007	&	 54316.53010     	&	15	&	60	&	75	&	14.24462	&	0.317	&	0.948	&	6.71583E+30	&	Fast	\\
04.09.2007	&	 54348.36803     	&	15	&	15	&	30	&	3.67867	&	0.286	&	1.162	&	1.73436E+30	&	Slow	\\
04.09.2007	&	 54348.37497     	&	15	&	30	&	45	&	5.97366	&	0.057	&	1.252	&	2.81637E+30	&	Slow	\\
04.09.2007	&	 54348.41749     	&	225	&	270	&	495	&	59.74276	&	0.243	&	1.090	&	2.81666E+31	&	Slow	\\
04.09.2007	&	 54348.42113     	&	45	&	180	&	225	&	27.28352	&	0.154	&	1.189	&	1.28632E+31	&	Fast	\\
04.09.2007	&	 54348.42339     	&	15	&	45	&	60	&	8.12777	&	0.246	&	1.108	&	3.83195E+30	&	Slow	\\
04.09.2007	&	 54348.42894     	&	435	&	1260	&	1695	&	245.28684	&	0.252	&	1.101	&	1.15644E+32	&	Slow	\\
04.09.2007	&	 54348.49088     	&	30	&	555	&	585	&	68.49528	&	0.353	&	1.029	&	3.22931E+31	&	Fast	\\
04.09.2007	&	 54348.53259     	&	105	&	105	&	210	&	22.31739	&	0.251	&	1.150	&	1.05219E+31	&	Slow	\\
04.09.2007	&	 54348.54624     	&	15	&	150	&	165	&	25.48638	&	0.327	&	1.043	&	1.20159E+31	&	Fast	\\
04.09.2007	&	 54348.57241     	&	45	&	45	&	90	&	20.44655	&	0.259	&	1.060	&	9.63983E+30	&	Slow	\\
13.09.2007	&	 54357.45391     	&	270	&	3202	&	3472	&	4260.46668	&	1.900	&	-0.324	&	2.00866E+33	&	Fast	\\
13.09.2007	&	 54357.50611     	&	90	&	525	&	615	&	164.36076	&	0.413	&	0.748	&	7.74903E+31	&	Fast	\\
13.09.2007	&	 54357.51479     	&	225	&	255	&	480	&	64.11248	&	0.221	&	1.058	&	3.02268E+31	&	Slow	\\
13.09.2007	&	 54357.51792     	&	15	&	30	&	45	&	4.89558	&	0.217	&	1.061	&	2.30809E+30	&	Slow	\\
13.09.2007	&	 54357.52603     	&	90	&	225	&	315	&	31.39548	&	0.195	&	1.105	&	1.48019E+31	&	Slow	\\
13.09.2007	&	 54357.53159     	&	45	&	45	&	90	&	31.21739	&	0.578	&	0.700	&	1.47179E+31	&	Slow	\\
13.09.2007	&	 54357.53367     	&	135	&	495	&	630	&	139.53195	&	0.320	&	0.942	&	6.57844E+31	&	Fast	\\
13.09.2007	&	 54357.54027     	&	15	&	90	&	105	&	16.08904	&	0.274	&	1.031	&	7.58541E+30	&	Fast	\\
13.09.2007	&	 54357.54391     	&	225	&	608	&	833	&	115.29866	&	0.259	&	1.021	&	5.43592E+31	&	Slow	\\
13.09.2007	&	 54357.55564     	&	375	&	330	&	705	&	86.03669	&	0.243	&	1.047	&	4.05633E+31	&	Slow	\\
13.09.2007	&	 54357.56033     	&	75	&	90	&	165	&	21.26797	&	0.218	&	1.091	&	1.00271E+31	&	Slow	\\
13.09.2007	&	 54357.56310     	&	150	&	90	&	240	&	28.46478	&	0.222	&	1.077	&	1.34201E+31	&	Slow	\\
13.09.2007	&	 54357.56467     	&	15	&	15	&	30	&	2.73049	&	0.230	&	1.060	&	1.28733E+30	&	Slow	\\
13.09.2007	&	 54357.56744     	&	195	&	345	&	540	&	51.13014	&	0.223	&	1.066	&	2.41060E+31	&	Slow	\\
13.09.2007	&	 54357.58622     	&	135	&	422	&	557	&	76.76394	&	0.267	&	0.958	&	3.61915E+31	&	Slow	\\
13.09.2007	&	 54357.59180     	&	60	&	45	&	105	&	14.50886	&	0.225	&	1.030	&	6.84041E+30	&	Slow	\\
13.09.2007	&	 54357.59492     	&	75	&	60	&	135	&	12.12054	&	0.238	&	1.031	&	5.71440E+30	&	Slow	\\
18.09.2007	&	 54362.30638     	&	12	&	12	&	24	&	4.85403	&	0.307	&	1.075	&	2.28850E+30	&	Slow	\\
18.09.2007	&	 54362.32015     	&	138	&	252	&	390	&	42.15909	&	0.273	&	1.190	&	1.98765E+31	&	Slow	\\
18.09.2007	&	 54362.53984     	&	96	&	60	&	156	&	25.40758	&	0.244	&	1.167	&	1.19788E+31	&	Slow	\\
18.09.2007	&	 54362.55264     	&	36	&	108	&	144	&	16.09805	&	0.289	&	1.133	&	7.58966E+30	&	Slow	\\
18.09.2007	&	 54362.55430     	&	36	&	36	&	72	&	6.24738	&	0.266	&	1.169	&	2.94542E+30	&	Slow	\\
18.09.2007	&	 54362.55597     	&	108	&	36	&	144	&	17.35319	&	0.306	&	1.134	&	8.18142E+30	&	Slow	\\
18.09.2007	&	 54362.56389     	&	648	&	538	&	1186	&	141.62994	&	0.180	&	1.241	&	6.67735E+31	&	Slow	\\
18.09.2007	&	 54362.57109     	&	24	&	36	&	60	&	7.93551	&	0.382	&	1.055	&	3.74131E+30	&	Slow	\\
18.09.2007	&	 54362.57386     	&	48	&	36	&	84	&	14.60747	&	0.303	&	1.138	&	6.88690E+30	&	Slow	\\
18.09.2007	&	 54362.60384     	&	96	&	185	&	281	&	41.67595	&	0.356	&	1.051	&	1.96487E+31	&	Slow	\\
23.09.2007	&	 54367.31400     	&	30	&	150	&	180	&	58.40560	&	0.624	&	0.947	&	2.75362E+31	&	Fast	\\
23.09.2007	&	 54367.31834     	&	15	&	15	&	30	&	10.36101	&	0.611	&	0.936	&	4.88485E+30	&	Slow	\\
\hline
\end{tabular}
\end{table*}

\setcounter{table}{2}
\begin{table*}
\setlength{\tabcolsep}{0.3 pt}
\centering
\caption{The parameters derived from analyses of the detected flares.\label{tbl-3}}
\begin{tabular}{@{}cccccccccc@{}}
\hline\hline
Observing	&	HJD of Flare & Rise &	Decay &	Total &	Equivalent	&	Flare	&	Flare	&	Flare	&	Flare	\\
 &	Maximum	& Time	&	Time &	Duration & Duration	& Amplitude	& U-B	& Energy	&		\\
Date	&	(+24 00000)	&	(s)	&	(s)	&	(s)	&	(s)	&	(mag)	&	(mag)	&	(erg)	&	Type  \\
\hline
23.09.2007	&	 54367.31956     	&	30	&	15	&	45	&	12.05673	&	0.587	&	1.019	&	5.68432E+30	&	Slow	\\
23.09.2007	&	 54367.32268     	&	45	&	45	&	90	&	31.64597	&	0.590	&	0.909	&	1.49200E+31	&	Slow	\\
23.09.2007	&	 54367.36543     	&	90	&	180	&	270	&	72.42159	&	0.454	&	0.991	&	3.41442E+31	&	Slow	\\
23.09.2007	&	 54367.36890     	&	90	&	90	&	180	&	57.38554	&	0.499	&	0.974	&	2.70552E+31	&	Slow	\\
23.09.2007	&	 54367.37872     	&	255	&	135	&	390	&	98.81383	&	0.545	&	0.919	&	4.65872E+31	&	Slow	\\
23.09.2007	&	 54367.38237     	&	75	&	270	&	345	&	91.03458	&	0.533	&	0.979	&	4.29196E+31	&	Fast	\\
23.09.2007	&	 54367.38741     	&	165	&	165	&	330	&	112.60634	&	0.554	&	1.005	&	5.30899E+31	&	Slow	\\
23.09.2007	&	 54367.39001     	&	60	&	90	&	150	&	43.89776	&	0.526	&	1.072	&	2.06962E+31	&	Slow	\\
23.09.2007	&	 54367.39348     	&	195	&	345	&	540	&	157.07515	&	0.516	&	0.988	&	7.40554E+31	&	Slow	\\
23.09.2007	&	 54367.39869     	&	105	&	315	&	420	&	161.72549	&	0.698	&	0.880	&	7.62478E+31	&	Slow	\\
23.09.2007	&	 54367.42463     	&	120	&	240	&	360	&	78.94735	&	0.372	&	1.099	&	3.72209E+31	&	Slow	\\
23.09.2007	&	 54367.43088     	&	75	&	150	&	225	&	56.78978	&	0.661	&	0.845	&	2.67744E+31	&	Slow	\\
23.09.2007	&	 54367.43366     	&	45	&	30	&	75	&	33.35158	&	0.630	&	0.874	&	1.57241E+31	&	Slow	\\
23.09.2007	&	 54367.43938     	&	30	&	60	&	90	&	21.49201	&	0.522	&	1.042	&	1.01327E+31	&	Slow	\\
23.09.2007	&	 54367.44095     	&	30	&	45	&	75	&	13.82339	&	0.513	&	1.129	&	6.51724E+30	&	Slow	\\
30.09.2007	&	 54374.46141     	&	75	&	495	&	570	&	117.72863	&	0.416	&	1.063	&	5.55049E+31	&	Fast	\\
30.09.2007	&	 54374.46888     	&	60	&	210	&	270	&	55.56658	&	0.428	&	1.047	&	2.61977E+31	&	Fast	\\
30.09.2007	&	 54374.47217     	&	75	&	255	&	330	&	76.63445	&	0.425	&	1.040	&	3.61304E+31	&	Fast	\\
10.10.2007	&	 54384.31223     	&	577	&	744	&	1321	&	482.01474	&	0.784	&	0.751	&	2.27253E+32	&	Slow	\\
10.10.2007	&	 54384.32514     	&	372	&	744	&	1116	&	288.43829	&	0.275	&	1.278	&	1.35988E+32	&	Slow	\\
10.10.2007	&	 54384.33889     	&	36	&	48	&	84	&	14.26053	&	0.462	&	1.119	&	6.72333E+30	&	Slow	\\
10.10.2007	&	 54384.34167     	&	36	&	24	&	60	&	21.88830	&	0.448	&	1.103	&	1.03196E+31	&	Slow	\\
10.10.2007	&	 54384.34709     	&	132	&	156	&	288	&	56.78634	&	0.347	&	1.181	&	2.67727E+31	&	Slow	\\
10.10.2007	&	 54384.35084     	&	168	&	204	&	372	&	59.84651	&	0.328	&	1.215	&	2.82155E+31	&	Slow	\\
10.10.2007	&	 54384.37766     	&	1057	&	1164	&	2221	&	617.18031	&	0.444	&	1.103	&	2.90979E+32	&	Slow	\\
10.10.2007	&	 54384.41952     	&	12	&	60	&	72	&	23.38733	&	0.498	&	1.027	&	1.10263E+31	&	Fast	\\
10.10.2007	&	 54384.46799     	&	24	&	12	&	36	&	7.28310	&	0.521	&	0.994	&	3.43372E+30	&	Slow	\\
10.10.2007	&	 54384.48174     	&	60	&	108	&	168	&	32.81632	&	0.603	&	0.881	&	1.54717E+31	&	Slow	\\
10.10.2007	&	 54384.49910     	&	24	&	84	&	108	&	31.69380	&	0.609	&	0.845	&	1.49425E+31	&	Fast	\\
10.10.2007	&	 54384.50133     	&	108	&	36	&	144	&	51.14498	&	0.564	&	0.941	&	2.41130E+31	&	Slow	\\
17.10.2007	&	 54391.25731     	&	48	&	540	&	588	&	72.54632	&	0.317	&	1.052	&	3.42030E+31	&	Fast	\\
17.10.2007	&	 54391.35890     	&	1164	&	2075	&	3239	&	418.76582	&	0.190	&	1.156	&	1.97433E+32	&	Slow	\\
\hline
\end{tabular}
\end{table*}

\begin{table*}
\begin{center}
\caption{For both fast and slow flares whose rise times are the same. The results obtained from both the regression calculations and the t-test analyses performed to the mean averages of the equivalent durations ($logP_{u}$) versus flare rise times ($logT_{r}$) in the logarithmic scale are listed.\label{tbl-4}}
\begin{tabular}{@{}lrr@{}}
\hline\hline					
\textbf{Flare Groups :} 	&	 \textbf{Slow Flare} 	&	 \textbf{Fast Flare} 	\\
\hline									
\textbf{\textit{Best Representation Values}}	&		&		\\				
Slope : 	&	1.046$\pm$0.048	&	1.232$\pm$0.181	\\
$y-intercept$ when $x = 0.0$ : 	&	-0.450$\pm$0.095	&	-0.137$\pm$0.302	\\
$x-intercept$ when $y = 0.0$ : 	&	0.430	&	0.111	\\
\hline\hline									
\textbf{\textit{Mean Average of All Y Values}} 	&	 	&	 	\\				
Mean Average : 	&	1.544	&	1.871	\\
Mean Average Error : 	&	0.067	&	0.130	\\
\hline\hline									
\textbf{\textit{Goodness of Fit}} 	&	 	&	 	\\				
$r^{2}$ : 	&	0.871	&	0.744	\\
\hline\hline									
\textbf{\textit{Is slope significantly non-zero?}} 	&	 	&	 	\\				
$p-value$ : 	&	 $<$ 0.0001 	&	$<$ 0.0001	\\
Deviation from zero? : 	&	 $Significant$ 	&	$Significant$	\\
\hline
\end{tabular}
\end{center}
\end{table*}

\begin{table*}
\begin{center}
\caption{Using the least-squares method, the parameters were obtained from the OPEA function.\label{tbl-5}}
\begin{tabular}{@{}lrr@{}}
\hline\hline
\textbf{Parameter}	&	\textbf{Value}	&	\textbf{Error}	\\
\hline					
$y_{0}$	&	0.972	&	0.057	\\
$plateau$	&	2.810	&	0.057	\\
$k$	&	0.001601	&	0.000258	\\
$tau$	&	624.8	&	-	\\
$half-life$	&	433.1	&	-	\\
$span$	&	1.837	&	0.115	\\
\hline
\end{tabular}
\end{center}
\end{table*}


\end{document}